\documentclass[a4paper,11pt,leqno]{amsart}

\makeatletter
\usepackage{hyperref}
\usepackage{esint,amssymb}
\usepackage{mathtools}
\usepackage[cyr]{aeguill}
\usepackage{lmodern}
\usepackage{bm}
\usepackage{amsfonts,dsfont,delarray,amssymb,amsmath,amsthm,a4,a4wide,color}
\usepackage[ansinew]{inputenc}

\newtheorem{theo}{Theorem}

\newtheorem{lem}{Lemma}
\newtheorem{coro}[theo]{Corollary}

\newtheorem{conj}{Conjecture}
\numberwithin{equation}{section}

\def\Xint#1{\mathchoice
   {\XXint\displaystyle\textstyle{#1}}%
   {\XXint\textstyle\scriptstyle{#1}}%
   {\XXint\scriptstyle\scriptscriptstyle{#1}}%
   {\XXint\scriptscriptstyle\scriptscriptstyle{#1}}%
   \!\int}
\def\XXint#1#2#3{{\setbox0=\hbox{$#1{#2#3}{\int}$}
     \vcenter{\hbox{$#2#3$}}\kern-.5\wd0}}
\def\dashint{\Xint-}

\def\1{\mathds{1}}

\def\div{\mathrm{div} \ }

\def\({\left(}
\def\){\right)}

\def \be{\begin{equation}}
\def \ee{\end{equation}}

\def\g{\mathsf{g}}

\def\nab{\nabla}
\def\yg{|y|^\gamma}
\def\R{\mathbb{R}}
\def\drd{\delta_{\R^\d}}
\def\f{\mathbf{f}}

\def\p{\partial}

\def\W{\mathbb{W}}
\def\d{\mathsf{d}}
\def\C{\mathcal{C}}

\title[Crystallization and the Cohn-Kumar conjecture]{Crystallization for Coulomb and Riesz interactions  as  a consequence of the Cohn-Kumar conjecture}
\author[M. Petrache]{Mircea Petrache}  
\address[M. Petrache]{PUC Chile, Facultad de Matematicas, Av. Vicuna Mackenna 4860, 6904441, Santiago, Chile} 
\email{decostruttivismo@gmail.com}

\author[S. Serfaty]{Sylvia Serfaty}
\address[S. Serfaty]{Courant Institute of Mathematical Sciences, New York University, 251 Mercer St., New York, NY 10012}
\email{serfaty@cims.nyu.edu}

\date{\today}

\begin{document}

\maketitle

\begin{abstract} The Cohn-Kumar conjecture states that the triangular lattice in dimension 2, the $E_8$ lattice in dimension 8, and the Leech lattice in dimension 24 are universally minimizing in the sense that they minimize the total pair interaction 
energy of infinite point configurations for all completely monotone functions of the squared distance. This conjecture was recently proved by Cohn-Kumar-Miller-Radchenko-Viazovska in dimensions 8 and 24. We explain in this note how the conjecture implies the minimality of the same lattices for the Coulomb and Riesz renormalized energies as well as jellium and  periodic jellium energies, hence settling the question of their minimization in dimensions 8 and 24.
 
\end{abstract}

\section{Introduction}
\subsection{The Cohn-Kumar conjecture and the main question}
 Define a point configuration $\C$ to be a nonempty, discrete, closed subset of Euclidean space $\R^\d$. For $p:\R_+\to \R$ any function, let the (lower) $p$-energy of $\C$ be 
\be \label{pener}  E_p(\C):= \liminf_{R \to \infty} \frac{1}{|\C \cap B_R|} \sum_{x, y \in \C \cap B_R, x\neq y}
p(|x-y|)\ee
where $B_R$ is the ball of center $0$ and radius $R$ in $\R^\d$.

We say that  $p$ is  a completely monotone function of the  squared distance when $p(r)= g(r^2)$
with $g$ a smooth completely monotone function on $\R_+$ i.e. satisfying $(-1)^k g^{(k)}(r) \ge 0$  for all $r \ge 0$ for every integer $k\ge 0$. This includes for instance Gaussians.

Let $\Lambda_0$ denote the triangular lattice $A_2$ in dimension $2$, the $E_8$ lattice in dimension $8$ and the Leech lattice in dimension $24$, dilated so that their fundamental cell has volume 1.
 We do not give here the precise definitions of the $E_8$ and Leech lattices, but suffice to say that these are Bravais lattices which means that they have the form $\sum_{i=1}^{\d} u_i \mathbb{Z}$ for some vectors $u_i \in \R^\d$, and that the triangular lattice in dimension 2 is the one spanned by two  vectors  of same norm forming an angle $\pi/3$.


\begin{conj}[{Cohn-Kumar~\cite{ck}}]
\label{conj.CK}
In dimension $\d=2,8$, resp. $24$, the lattice $\Lambda_0$ is {\it universally minimizing} in the sense that it
  minimizes $E_p$ among all possible point configurations  of  density $1$  for all $p$'s that are  completely monotone functions of the squared distance.\end{conj}
  Coulangeon and Sch\"urmann proved in \cite{cs} a local version of this conjecture with the result that $\Lambda_0$ is a {\it local} minimizer of $E_p$. Then
Conjecture~\ref{conj.CK} was recently proved in dimensions~8 and~24 --- it remains open in dimension~2.

\begin{theo}[{Cohn-Kumar-Miller-Radchenko-Viazovska~\cite{5}}] 
\label{thck}
The Cohn-Kumar conjecture is true in dimensions $\d=8$ and $24$.
\end{theo}

This breakthrough result was itself made possible by the seminal works of Viazovska~\cite{via} and the same authors~\cite{ckrmv} on the solution of the sphere packing problem in the same dimensions (see also the expository papers~\cite{cohn,dlv}). As one of very few proofs of crystallization in dimensions larger than one, it represents major progress on the topic. Indeed, results stating that optimal configurations for an interaction energy are periodic, much less lattices, are extremely rare (for more on this, see the review \cite{blanclewin}):  besides the solutions to sphere packing problems in dimensions 2 and 3, another famous result is that of \cite{radin,theil} for a particular short range interaction in dimension 2, perturbing off of the sphere packing problem. Note that, without the complete monotonicity assumption, the triangular lattice cannot be expected to always be the minimizer for such problems ; for instance  \cite{bpl} recently gave an example of a nonmonotone interaction function for which the minimizer can be proven to be {\it square} lattice instead of the triangular one.

The goal of this note is to explore consequences of this result (or, in the case of dimension~2, of the Cohn-Kumar conjecture) for particular nonsmooth interactions, more specifically Coulomb and Riesz interactions, which are of particular importance in physics and in approximation theory. 

\smallskip
 
We now restrict to interactions  of the form~$p(r)=r^{-s}$, or~$p(r)=-\log r$ in dimension~2. If~$s>\d$ then the~$p$-energy \eqref{pener} is typically (for good configurations) well summable, and writing $r^{-s}$ as a superposition of Gaussians
\be\label{suprs}
\frac{1}{r^s} = \frac{1}{\Gamma(s/2)}\int_0^\infty e^{-t r^2} t^{\frac{s}{2}-1}dt,\ee
it follows immediately that the Cohn-Kumar conjecture for smooth functions (or just Gaussians) implies the Cohn-Kumar conjecture for $p(r)= r^{-s}$, i.e. the optimality of $\Lambda_0$ for $E_p$.

In contrast, when $p(r)= r^{-s}$ with  $s<\d$, the energy $E_{p}$ is always infinite because of the divergence of the series of pair interactions. This difficulty is due to the long-range nature of the interaction, and includes the very important case of Coulomb interactions, corresponding to $s=\d-2$ in dimension $\d\ge 3$ and $-\log r$  in dimension 2 \footnote{we take the convention $s=0$ to denote  the logarithmic case}. It is in fact not straightforward to give a good definition of total Coulomb or Riesz interaction for general infinite point  configurations, we next explore this question.

\subsection{Coulomb and Riesz interactions: definitions and motivations}
Let 
\be\label{defg} \g(x):=|x|^{-s},\ee or $\g(x)=-\log |x|$ in dimension $\d=2$,
 with $\max(\d-2,0)\le s <\d$.  
A definition of total $\g$-energy was proposed  first in   \cite{ssgl} for the $\d=2$ logarithmic case, then in \cite{rs} for the  Coulomb case in any dimension $\d\ge 2$, and then in \cite{ps} for the Riesz interactions with $\max(\d-2,0)\le s <\d$ in any dimension.
It consists in defining the interaction for the infinite configuration of points (say of density $1$) screened or  neutralized by a uniform background of charge $-1$, via the Coulomb or Riesz potential generated by $\sum_{x\in \C} \delta_x-1$.  
 Because of the divergence of the Coulomb (or Riesz) potential at each point, the energy needed to be defined in a ``renormalized way", hence it was called ``renormalized energy", also by analogy with the work of \cite{bbh}. The precise definition of this energy $\W$ is recalled in Appendix \ref{app}.
 In physics,  such a neutralized system is called a {\it jellium} and was first introduced by Wigner \cite{wigner} (who was really focusing on the quantum case) who conjectured that the minimum should be achieved by a crystal, now called a ``Wigner crystal", in dimensions 1, 2 and 3. More precisely, the jellium minimization problem  is usually stated as the question of minimizing  
 \be \label{jellium} 
 \mathrm{e}_{\mathrm{Jel}}:=\lim_{R\to \infty} \frac{1}{R^\d}\( \min_{a_1, \dots, a_N\in K_R} 
 \sum_{i \neq j} \g(a_i-a_j) -2 \sum_{i=1}^N \int_{K_R} \g(a_i-y) dy+\iint_{K_R\times K_R}\g(x-y) dxdy\)\ee
 with $K_R= [-\frac{R}{2}, \frac{R}{2}]^\d$ and $N=R^\d$, where we have chosen for instance the fundamental cell to be a cube, but any nondegenerate shape can be used. The existence of the limit in this definition was first proven in \cite{liebnarn}. In contrast  with the  function  $\W$ of \cite{ssgl,rs,ps}, \eqref{jellium} defines a minimization problem  but not an energy for arbitrary infinite configurations of points.

Also in contrast with \eqref{jellium}, in the  reformulation of \cite{ssgl,rs,ps} it is crucial that the Coulomb kernel is the fundamental solution of the Laplacian. Similarly, the Riesz kernel for $\max(\d-2,0)\le s<\d$ is the fundamental solution to the fractional Laplacian $(-\Delta)^{\alpha}$ with $\alpha= \frac{\d-s}{2}$, which can be turned into a local operator by the extension procedure of \cite{caffsilvestre}. This is the reason for the limitation to $s \ge \d-2$  in those works (in contrast, $s\le \d-2$ does not correspond to a $(-\Delta)^{\alpha}$ but to a higher order operator).

Another possible approach is to restrict to periodic configurations -- in the Coulomb case, this is also called the  {\it periodic jellium}.  The fact that $\g$ defined in \eqref{defg} 
is the fundamental solution for the Laplacian or for a fractional Laplacian then makes it possible to use for the definition the Green's function of a torus or the equivalent notion for the fractional Laplacian. 
More precisely, let $\Lambda $ be a lattice in $\R^\d$  of covolume 1 and $n$ an integer,  $N=n^\d$, and let $G_{n\Lambda} $ 
solve
  \begin{equation}\label{defG}
   (-\Delta)^\alpha G_{n\Lambda}=\delta_0 - \frac{1}{N} \quad \text{in } \ \R^\d/ (n\Lambda),  \qquad  \int_{R^\d/(n\Lambda)} G_{n\Lambda}=0,
  \end{equation}
  with \be\label{defalpha} \alpha= \frac{\d-s}{2}.\ee
  Here the fractional Laplacian can for instance be defined as corresponding to the Fourier multiplier $|\xi|^{2\alpha}$. The Fourier series expansion of $G_{n\Lambda}$ is  then
  \be  G_{n\Lambda}(x)= \frac{1}{N}\sum_{w \in \Lambda^* \backslash \{0\}} \frac{e^{2i\pi w \cdot x}}{(2\pi |w|)^{2\alpha} }. \ee

  For a configuration of $N$ points $a_i$ in $\R^\d/(n\Lambda)$ we may now define the periodic Riesz interaction energy
\be \label{2defW}
\mathcal W_{n\Lambda}(a_i, \dots, a_N)= c_{\d, s}^2 \(\frac{1}{N} \sum_{i\neq j}G_{n\Lambda} (a_i-a_j) + M_{n\Lambda}\),\ee
where for any lattice $\Lambda$, $M_\Lambda$ is the so-called Madelung constant of the lattice, defined by 
\be M_\Lambda:= \lim_{x\to 0 } (G_{\Lambda}(x) - c_{\d, s} \g(x))\ee and $c_{\d,s}$ is the constant defined by $(-\Delta)^{\alpha} \g=c_{\d, s} \delta_0$.

One can  easily check that  for $n\Lambda$-periodic configurations of the form $\cup_{i=1}^N \{a_i + n\Lambda\}$,  $\W$ coincides with $\mathcal W_{n \Lambda}$.
Another route for defining the energy of non-periodic infinite configurations is then to try to extend the formula \eqref{2defW}, such an approach was initiated in \cite{bs} for point processes, and the comparison of the various definitions was explored in more details in \cite{leble}. The non-periodic situation is in fact quite subtle as illustrated in    \cite{gesandier} who   also provide  necessary conditions and sufficient conditions for a configuration to have finite $\W$ energy in terms of the growth rate of its discrepancy in balls.

In \cite{ssgl} it was shown that  $\W$ can be derived as the limiting interaction energy for vortices in the 2D Ginzburg-Landau model of superconductivity. The same was done starting from the Ohta-Kawasaki model (essentially a two-dimensional  version of Gamow's liquid drop model for matter) in \cite{gms}.
 In \cite{ss2,rs,ps}, $\W$ was derived as the limiting (as $N \to \infty$) interaction energy for (scaling limits of) minimizers of Coulomb and Riesz energies at finite number $N$ of points (with a confining potential). In \cite{lebles}, $\W$ was shown to also govern the random point configurations obtained as limits of Coulomb and Riesz gases (i.e. typical configurations under the Gibbs measure, at  inverse temperature $\beta$). For more background, in particular on the physical aspects, we refer to \cite{sicm,ln}.
These derivations in fact  originally {\it naturally motivated} the definition of $\W$ discussed above and showed that it is relevant, and they also reduced the corresponding initial problems to the question of minimization of $\W$.

In dimension 2,  $\W$ was shown to be minimized  within the class of lattices of covolume 1 by the triangular lattice in \cite{ssgl}, respectively \cite{ps}, by showing that this question could be  reduced to the question of minimizing the Epstein zeta function, previously solved in \cite{cassels,rankin,ennola,ennola2,diananda,montgomery} (see also \cite{chiu,sarnak,osp} for general dimension).
In \cite{ssgl} it was conjectured, in view of the observations of triangular {\it Abrikosov lattices} of vortices in superconductors (and  in line with the Cohn-Kumar conjecture) that the triangular lattice should minimize $\W$ in the 2D logarithmic case, among the class of all configurations of density~$1$. This conjecture is still open, but we show here that it is implied by the Cohn-Kumar conjecture.  Related to that physical motivation, let us point out that the triangular Abrikosov lattice also arises in a different regime of higher magnetic fields in the Ginzburg-Landau model as well as in superfluids model, the minimization problem is then  different and consists in minimizing of the fourth power of a product of Theta functions (see for instance \cite{abn,aftserf,nier}) which does not obviously seem to be of pair interaction type --- it would be very interesting to see if the Cohn-Kumar conjecture could help for that question too. 

\medskip

In \cite{ssgl,rs,ps}  (this was also detailed in \cite{cp})  it was shown 
 by a (rather delicate) {\it screening procedure} that for any $\Lambda$ and any $ \max(\d-2,0) \le s<\d$, 
 there exists a minimizing sequence for $\W$  formed of $n\Lambda$-periodic configurations
with period $n\to \infty$, hence  for every $\max(\d-2,0)\le s<\d$ and every lattice $\Lambda$ of covolume 1, 
 \be \label{egene}\min \W= \lim_{n \to \infty}  \min \mathcal W_{n\Lambda} 
 .\ee
    One can also show with the same ideas that,  up to multiplicative and additive constants
    $$\mathrm{e}_{\mathrm{Jel}}=\lim_{n \to \infty}  \min \mathcal W_{n\Lambda}. $$
   This is one of the key steps in the proof of the main result of \cite{cp},   and an alternative short proof  is  provided in \cite{lls} in the Coulomb case.
   
 The screening procedure also allowed to prove in \cite{rns,PRN} a result of ``equidistribution" of the points and the energy of minimizers, in the Coulomb case, which is still weaker than a periodicity result. 
 
   As a result of    \eqref{egene}, in order to identify $\min \W$  we can reduce to computing $\lim_{n\to \infty}\min \mathcal{W}_{n \Lambda_0}$. This reduction  is the most delicate and longest step in the non-Coulomb case, as the final result is subsequently an easy consequence of the following formula
   \begin{lem}\label{lem1} Let $\Lambda$ be a lattice of covolume $1$, $n$ an integer, and $N=n^\d$. The function  $G_{n\Lambda}$ being as in \eqref{defG}, 
we have 
\be \label{Gint}
G_{n\Lambda} (x)=
\frac{1}{\Gamma(\frac{\d-s}{2})} \int_0^\infty \(  \sum_{ v \in n\Lambda}\Psi_t(x-v) -\frac{1}{N}\) t^{\frac{\d-s}{2}-1} \, dt\ee
 where $\Psi_t(x)$ is the standard heat kernel $(4\pi t)^{-\frac\d2} e^{-|x|^2/(4t)}$.\end{lem}
 
 We will see in Section \ref{2.1} that  for any $x \notin  n\Lambda$, this integral makes sense for any $0\le s<\d$, which gives a way of extending the definition of $\mathcal W_{n\Lambda}$ to all $0\le s<\d$. The formula \eqref{Gint} for $\Lambda = \mathbb Z^d$ appears for instance  in \cite{roncalstinga}. 
 
Formula \eqref{Gint} is the central point of the proof. It expresses the periodic Riesz kernel $G_{n\Lambda}$ as the Mellin transform of the 
heat kernel on the torus $  \sum_{ v \in n\Lambda}\Psi_t(x-v) -\frac{1}{N}$, except extended to $\d-s<\d$ . 
 
Since the standard heat kernel $\Psi_t$  is a completely monotone function of the squared distance, the Cohn-Kumar conjecture applies to it and  the same then holds for $G_{n\Lambda}$ by integration. This allows to identify the minimum of $\mathcal W_{n\Lambda_0}$, and then that of  $\W$ by taking the large $n$ limit (details are in Section \ref{2.3}).
 \medskip
 
 Yet another point of view on the question of summing Coulomb interactions is that of Ewald summation (essentially, an instance of Poisson summation). It consists in  defining a periodic Riesz function as the analytic continuation of the Epstein-Hurwitz zeta function
\be \zeta_\Lambda(s,x):= \sum_{v\in \Lambda} \frac{1}{|x+v|^s}\ee
naturally defined for $s>\d$ and $x\notin \Lambda$.
This is made rigorous in  \cite{hss,hsss} (see also \cite[Chap. 10]{bhsbook}) via  ``convergence factors" (on the physics side, see \cite{borwein} and references therein).
For $s>\d$ one can prove that 
\be \zeta_\Lambda(s,x)+ \frac{2\pi^{\frac\d2}}{|\Lambda|\Gamma(\frac{s}{2})(\d-s)}=  
F_{s, \Lambda}(x)
\ee
with 
\be \label{defF} F_{s, \Lambda}(x):= \sum_{v\in \Lambda} \int_1^\infty e^{-|x+v|^2 t}\frac{t^{\frac{s}{2}-1}}{\Gamma(\frac{s}{2})} dt
+ \frac{1}{|\Lambda|}\sum_{w\in \Lambda^* \backslash \{0\}}e^{2i\pi w\cdot x}\int_0^1 \frac{\pi^{\frac\d2}}{t^{\frac\d2}}e^{-\frac{\pi^2 |w|^2}{t} }\frac{t^{\frac{s}{2}-1}}{\Gamma(\frac{s}{2})} dt\ee  (this is similar to the calculations made by Riemann on his zeta function)
hence $F_{s, \Lambda} (x)- \frac{2\pi^{\frac\d2}}{|\Lambda|\Gamma(\frac{s}{2})(\d-s)}$ provides an analytic continuation of $\zeta_\Lambda(\cdot, x)$ and thus a definition for any $s$. 
The authors of \cite{hss,hsss,bhsbook} then use $F_{s, \Lambda}$ as the periodized Riesz interaction function of the torus $\R^\d/\Lambda$.
Starting from \eqref{Gint}, decomposing the integral into the integral over $(0,1)$  and that over $(1,\infty)$ and using the Poisson summation formula for the first part (see details in Section \ref{2.2}), one checks that $F_{s, \Lambda}$  is (up to constants) equal to $G_{\Lambda}$, so the same result as below applies to $\sum_{i\neq j} F_{s,n\Lambda_0} (a_i-a_j)$.
Note that a similar formula to \eqref{Gint} but relating the Epstein zeta function to the Jacobi Theta function is also well-known, see for instance \cite[p. 8]{5} and  \cite[Chap. 10]{bhsbook}, and this formula can be used to retrieve \eqref{Gint}.

\subsection{Result}

We can now provide a complete statement.
\begin{theo}
\label{t.theorem}
If the Cohn-Kumar conjecture (for smooth functions) is true, then 
 $\Lambda_0$ achieves the minimum  of  $\mathcal W_{n \Lambda_0}$ for every integer $n$ and every $0\le s<\d$, hence of $ \mathrm{e}_{\mathrm{Jel}}$ and of  $\W$ in all Coulomb cases ($\d\ge 2$) and all Riesz cases with $\max(\d-2,0)\le s<\d$. 
 \end{theo}
Of course, since the conjecture holds in dimension 8 and 24 we then have a positive answer:
\begin{coro}
The $E_8$ lattice in dimension 8, resp. Leech lattice in dimension 24, achieves the minima above. \end{coro}
Combined with the results  of \cite{rs,ps}, this shows a complete crystallization result at zero temperature  for the Coulomb and Riesz gases in dimensions 8 and 24.

As mentioned above, in dimension 2 and for the Coulomb interaction, it was conjectured in \cite{ssgl} that the triangular lattice achieves the minimum of $\W$. By analyzing the minimization of the logarithmic energy on the 2-sphere, it was then shown by B\'etermin and Sandier \cite{bs} that this conjecture is equivalent to a conjecture of Brauchart-Hardin-Saff made in \cite{bhs} by  an analytic continuation argument.
We can thus say that
\begin{coro}
In dimension 2, the Cohn-Kumar conjecture implies the mutually equivalent conjectures of \cite{bhs}, \cite{ssgl} and \cite{wigner}.
\end{coro}
Thus,  proving the Cohn-Kumar conjecture in dimension 2 would prove the Wigner conjecture in dimension 2, complete the program of  \cite{ssgl} of proving the emergence of the triangular Abrikosov lattice of vortices starting from the Ginzburg-Landau model of superconductivity,  as well as  complete the proof of crystallization of the two-dimensional Coulomb gas at zero temperature in \cite{ss2}. 

\medskip

We  have not said anything about the uniqueness of $\Lambda_0$ as a minimizer.  Because of the definition \eqref{pener}, uniqueness cannot hold since any finite set  perturbation of $\Lambda_0$ remains a minimizer. The same is true for $\W$ 
in view of its definition \eqref{defWW}--\eqref{defWW2}. However, when restricting to a periodic situation and minimizing $\mathcal W_{n\Lambda_0}$ for fixed $n$, the result of \cite{5} gives uniqueness of $\Lambda_0$ up to isometries. 

\medskip

{\bf Acknowledgements: } MP is supported
by the Fondecyt Iniciaci\'on grant number 11170264 entitled ``Sharp asymptotics for large particle systems and
topological singularities''. SS is supported by by NSF grant DMS-1700278 and by a Simons Investigator grant. 
She wishes to thank Stephen Miller for stimulating the writing of this note and to him, Mathieu Lewin and Ed Saff for helpful comments on the first draft.
\section{Proofs}
\subsection{Proof of Lemma \ref{lem1} }\label{2.1}

First, we note that $$\sum_{v\in n\Lambda} \Psi_t(x-v)-\frac1N:= \Phi_t(x) $$ 
is the heat kernel of $\R^\d/(n\Lambda)$ (with average $0$). By the spectral gap of the torus, it decays exponentially fast in time, hence the integral in \eqref{Gint} converges.

For the usual Coulomb kernel (up to a multiplicative constant) $\g$  with $s=\d-2$ (or $\g=-\log $ in dimension 2) it is well-known that at least in dimension $\d\ge 3$, 
$\mathsf{g}(x)$ is the integral in time of the standard heat kernel. 
To get the similar
formula with weight \eqref{Gint} for the fractional periodic Green function in all dimensions we can either proceed by Fourier transform and using the formula \eqref{suprs}, or by spectral representation of $(-\Delta)$ and using \eqref{suprs}, as follows.

\medskip

The operator $(-\Delta)^{-\frac{\mathsf{d}-s}2}$ is well-defined over the $L^2$-orthogonal space to the constant functions on $\mathbb R^d/\Lambda$. If $0=\lambda_1<\lambda_2\le \cdots$ are the eigenvalues of $-\Delta$ over $L^2(\mathbb R^d/\Lambda)$ counted with multiplicity and $\varphi_k$ are corresponding eigenfunctions, then
\[
 G_\Lambda(x-y)=\sum_{k\ge 2} \lambda_k^{-\frac{\mathsf{d}-s}2}\varphi_k(x)\varphi_k(y)\quad\mbox{and}\quad \Psi_t(x-y)=\sum_{k\ge 1} e^{-t\lambda_k}\varphi_k(x)\varphi_k(y).
\]
The $L^2$-projection onto the constant functions (i.e. onto the eigenspace corresponding to $\lambda_1$) is given by the constant kernel $N^{-1}$. 
Now using the formula $\lambda_k^{-\alpha}=\frac{1}{\Gamma(\alpha)}\int_0^\infty e^{-t\lambda_k}t^{\alpha-1}dt$ for $\alpha=\frac{\mathsf{d}-s}2$ and by the orthogonality of the $\varphi_k$, we find the desired formula \eqref{Gint}.

\subsection{Equality between $F_{s, \Lambda} $ and $G_{\Lambda}$}\label{2.2} Starting from the definition of $G_\Lambda$ in \eqref{Gint}, we  split the integral  into two intervals, $[0,1]$ and $[1, \infty)$. In the first interval we rewrite 
$$\sum_{v\in \Lambda}   \Psi_t(x-v) - \frac{1}{|\Lambda|} = \Phi_t (x) = \sum_{v\in \Lambda} \frac{e^{- \frac{|x+v|^2}{4t}}}{ (4\pi t)^{\frac\d2}} - \frac{1}{|\Lambda|}.$$
In the second interval, we may use  Poisson summation to rewrite 
$$   \Phi_t(x)= \frac{1}{|\Lambda|} \sum_{w\in \Lambda^* \backslash \{0\}} e^{-4\pi^2 |w|^2 t } e^{2i\pi w\cdot x}.$$
We then integrate over each interval and perform the change of variables $t \to 1/t$ to retrieve  $F_{s, \Lambda}$ as in \eqref{defF}, up to multiplicative  and additive constants.

\subsection{Proof of Theorem~\ref{t.theorem}} \label{2.3}
 If   $a_1, \dots, a_N$ is a configuration of $N$ points in $\R^\d/(n\Lambda)$, then the $n\Lambda$-periodic configuration   in the whole $\R^\d$ formed by  $\cup_{i=1}^N\{ a_i + n\Lambda\}$ has $p$-energy equal to 
$$
\frac{1}{N}\sum_{j\neq k} \sum_{v\in n\Lambda \backslash \{a_k-a_j\}} p \( |v+ a_j-a_k|\).$$
If the Cohn-Kumar conjecture holds, then this quantity must be larger than that of $\Lambda_0$. We deduce that, for $p$ smooth and completely monotone in the squared distance, we have
\be \label{minp}
\frac{1}{N}\sum_{j\neq k} \sum_{v\in n\Lambda \backslash \{a_k-a_j\}} p \( |v+ a_j-a_k|\)\ge E_p(\Lambda_0)
= \frac{1}{N}\sum_{j\neq k}\sum_{v\in n\Lambda_0 \backslash \{a_k^0 -a_j^0\}} p \( |v+ a_j^0-a_k^0|\)
 .\ee
Here we view $a_j^0$ as the configuration of points of $\Lambda_0 $ in $\R^\d/(n\Lambda_0)$ which is also identified informally with $\Lambda_0$ itself. 

Applying \eqref{minp} to $p=\Psi_t$ (which is completely monotone in the squared distance), and plugging into \eqref{Gint} we deduce that, for any configuration $a_1, \dots, a_N$ in $\R^\d/ (n\Lambda_0)$, 
\begin{align*}
\sum_{j\neq k} G_{n\Lambda_0} (a_j-a_k) 
 &
=
\frac{1}{\Gamma(\frac{\d-s}{2})} \sum_{j\neq k}  \int_0^\infty \( \sum_{ v \in n \Lambda_0}\Psi_t(a_j-a_k-v) -\frac{1}{N}\)t^{\frac{\d-s}{2}-1}  \, dt
\\ & 
\geq 
\frac{1}{\Gamma(\frac{\d-s}{2}) }\sum_{j\neq k} \int_0^\infty  \( \sum_{ v \in n \Lambda_0} \Psi_t(a_j^0-a_k^0-v) -\frac{1}{N}\) t^{\frac{\d-s}{2}-1}\, dt
\\ & 
=  \sum_{j\neq k} G_{n\Lambda_0} (a_j^0-a_k^0) .
\end{align*} In view of \eqref{2defW}
 and \eqref{egene} this completes the proof.

\appendix

\section{Definition of $\W$}\label{app}
For the interested reader we recall  here the full definition of $\W$  from \cite{ssgl,rs,ps}, with simplifications from \cite{lebles}.
Let us start with the Coulomb case which is the easiest. 

Let $K_R$ denote the cube $[-R/2,R/2]^\d$ and $\dashint$ the average. 
For any $\eta \in (0,1)$, we define
\begin{equation}
\label{def:truncation} \g_{\eta} := \min(\g, \g(\eta)), \quad \f_{\eta} := \g - \g_{\eta}.
\end{equation}

Given  a function $h$ corresponding to a Coulomb potential generated by $\C-1$, that is satisfying a relation of the form 
$$-\Delta h= c_\d \(  \sum_{x\in \C} \delta_x-1\)$$
we set 
\be \label{heta} h_\eta:= h-\sum_{x\in \C} \nab \f_\eta(\cdot -x)\ee
which corresponds to the same potential but truncated near each $x\in \C$ (or equivalently with charges spread over $\p B(x, \eta)$). 
Given an infinite point configuration in $\R^\d$ the Coulomb renormalized energy is defined as 
\be \label{defWW}\W(\C)= \inf\left\{ \liminf_{R_\to \infty}  
\lim_{\eta \to 0}\dashint_{K_R} |\nab h_\eta|^2 - c_\d \g(\eta), \quad - \Delta h= c_\d \( \sum_{x\in \C} \delta_x-1\) \ \text{in } \ \R^\d\right\}
.\ee

The Riesz case is a little more complicated since then the Riesz kernel is not the kernel of a local operator. To handle this, we use the Caffarelli-Silvestre extension procedure \cite{caffsilvestre}: we work in the extended space $\R^{\d}\times \R$ or $\R^{\d+1}$ (and identify $\R^\d$ with $\R^\d \times \{0\})$ with the last variable denoted $y$. We let $\drd$ denote the uniform measure on $\R^\d \times \{0\}$. Then $\g(x)=|x|^{-s}$ is the kernel of the operator $-\div (\yg \nab \cdot )$ in $\R^{\d+1}$ for $\gamma= s-\d+1$.
We still use the definition \eqref{def:truncation} as well as \eqref{heta}.

The Riesz 
renormalized energy of $\C$  is then defined by 
\begin{multline}\label{defWW2} \W(\C)= \inf\Bigg\{ \liminf_{R_\to \infty}  
\lim_{\eta \to 0}\frac{1}{R^\d} \int_{K_R\times \R}\yg |\nab h_\eta|^2 - c_{\d,s} \g(\eta), \quad\\
 - \div (\yg\nab  h)= c_{\d,s} \( \sum_{x\in \C} \delta_{(x,0)}-\drd\) \ \text{in } \ \R^{\d+1}\Bigg\}
.\end{multline}


\begin{thebibliography}{99}
\bibitem[AS]{aftserf}
A.
Aftalion, S. Serfaty, Lowest Landau level approach for the
Abrikosov lattice  of a superconductor close to $H_{c_2}$, 
 {\it Selecta Math, 13}, (2007), No 2, 183-202.
 
 \bibitem[ABN]{abn} 
A. Aftalion, X. Blanc, F. Nier, Lowest Landau level functional and Bargmann spaces for Bose-Einstein condensates, {\it J. Funct. Anal.} {\bf  241}  (2006), 661--702.
 
\bibitem[BBH]{bbh}
F. Bethuel, H. Brezis, F. H\'elein, {\it Ginzburg-Landau Vortices}, Birkh\"auser, 1994.


\bibitem[BPL]{bpl} L. B\'etermin, M. Petrache, L. de Luca,  Crystallization to a square lattice for a 2-body potential, 
{\tt arXiv:1907.06105.}
\bibitem[BS]{bs} L. B\'etermin, E. Sandier, Renormalized energy and asymptotic expansion of optimal logarithmic energy on the sphere,  {\it  Constr. Approx.} {\bf 47}, No 1, (2018), 39-74.



\bibitem[BLe]{blanclewin} X. Blanc, M. Lewin, The Crystallization Conjecture: A Review, {\it EMS surveys} {\bf 2} (2015), 255--306.

\bibitem[BoHS]{bhsbook}  S. Borodachov, D. Hardin, E. Saff,
{\it Discrete Energy on Rectifiable Sets}, to appear in Springer  Mathematical Monographs.

\bibitem[BoS]{bos} A. Borodin, S. Serfaty, Renormalized Energy  Concentration in Random Matrices,  {\it  Comm. Math. Phys.} {\bf  320}, No 1, (2013), 199--244.

\bibitem[BGMWZ]{borwein}
J. Borwein,  M. Glasser, R. McPhedran,  J. Wan,  I.  Zucker,  {\it Lattice sums, then and now}, Cambridge University Press, 2013.

\bibitem[BrHS]{bhs} S. Brauchart, D. Hardin, E. Saff, The next-order term for optimal Riesz and logarithmic energy asymptotics on the sphere. {\it Recent advances in orthogonal polynomials, special functions, and their applications,} 31-61, Contemp. Math., 578, Amer. Math. Soc., Providence, RI, 2012. 


\bibitem[CS]{caffsilvestre} L. A. Caffarelli, L. Silvestre, An extension problem related to the fractional Laplacian, {\it Comm. PDE} {\bf 32}, (2007), no 7-9, 1245--1260.

\bibitem[Cas]{cassels} J. W. S. Cassels, On a problem of Rankin about the
Epstein zeta-function. {\it Proc. Glasgow Math. Assoc.} { \bf  4}  (1959), 73--80.

\bibitem[Ch]{chiu} P. Chiu, Height of flat tori, {\it Proc. Amer. Math. Soc.} {\bf 125} (1997), no. 3, 723--730.

\bibitem[Coh]{cohn} H. Cohn, A conceptual breakthrough in sphere packing,  {\it Notices AMS} {\bf 64} (2017), no. 2, 102--115.


\bibitem[CK]{ck} H.  Cohn, A.  Kumar,  Universally optimal distribution of points on spheres.  {\it J. Amer. Math. Soc. } {\bf  20}  (2007),  no. 1, 99--148.

\bibitem[CKMRV1]{ckrmv} H. Cohn, A. Kumar, S. D. Miller, D. Radchenko, M. Viazovska,
The sphere packing problem in dimension 24, {\it Ann. of Math.} (2) {\bf 185} (2017), no. 3, 1017--1033.

\bibitem[CKMRV2]{5} H. Cohn, A. Kumar, S. D. Miller, D. Radchenko, M. Viazovska, Universal optimality of the E8 and Leech lattices and interpolation formulas, {\tt arXiv:1902.05438.}

\bibitem[CP]{cp} C. Cotar, M. Petrache, Equality of the Jellium and Uniform Electron Gas next-order asymptotic terms for Coulomb and Riesz potentials,  {\tt arXiv:1707.07664.}

\bibitem[CS]{cs} R. Coulangeon, A. Sch\"urmann, 
Energy minimization, periodic sets and spherical designs, {\it Int. Math. Res. Not.}  (2012), no. 4, 829--848. 
\bibitem[Dia]{diananda}
P. H. Diananda, Notes on two lemmas concerning the Epstein zeta-function. {\it  Proc. Glasgow Math. Assoc. }  {\bf 6},   (1964),  202--204.

\bibitem[Enno1]{ennola} V.  Ennola, A remark about the Epstein zeta function.  {\it Proc. Glasgow Math. Assoc. } {\bf 6}, (1964), 198--201.


\bibitem[Enno2]{ennola2}
V. Ennola,  On a problem about the Epstein zeta-function.
 {\it  Proc. Cambridge Philos. Soc.} {\bf  60},  855--875,
(1964).

\bibitem[GS]{gesandier}Y. Ge, E. Sandier,
On lattices with finite Coulombian interaction energy in the plane, {\tt arXiv:1307.2621}.

\bibitem[GMS]{gms}
D. Goldman, C. Muratov, S. Serfaty, The Gamma-limit of the two-dimensional Ohta-Kawasaki functional. Part II: Droplet arrangement at the sharp-interface level via the Renormalized Energy,  {\it Arch. Rat. Mech. Anal. }  {\bf 212} (2014), no. 2, 445--501.

\bibitem[HSS]{hss} D. Hardin, E. Saff, B. Simanek,  Periodic discrete energy for long-range potentials, {\it  J.
Math. Phys.} {\bf 55}  (2014), no. 12, 123509.
 
 \bibitem[HSSS]{hsss} D. Hardin, E. Saff, B. Simanek, Y. Su, Next order energy asymptotics for Riesz potentials on flat tori, 
{\it Inter. Math. Res. Not.}  (2017), No. 12, 3529--3556.

\bibitem[dLV]{dlv} D. de Laat, F. Vallentin, A breakthrough in sphere packing : the search for magic functions, {\it Nieuw Archief voor Wiskunde} (5) {\bf 17} (2016), 184--192.
 \bibitem[Le]{leble} T. Lebl\'e, Logarithmic, Coulomb and Riesz energy of point processes, {\it J. Stat. Phys} {\bf 162} (4), (2016),  887--923. 
 
\bibitem[LS1]{lebles} T. Lebl\'e, S. Serfaty,   Large {D}eviation {P}rinciple for Empirical Fields of {L}og and {R}iesz gases, {\it Inventiones Math.} {\bf 210}, No. 3,  645--757.

\bibitem[LLS]{lls} M. Lewin, E. Lieb, R. Seiringer,
A floating Wigner crystal with no boundary charge fluctuations, to appear in {\it Phys. Rev. B}.

\bibitem[LN]{liebnarn} E. Lieb, H. Narnhofer, The thermodynamic limit for jellium, {\it J. Stat. Phys.}  {\bf 12}, No 4 (1975),  291--310.

\bibitem[Mont]{montgomery} H. L.  Montgomery,
  Minimal Theta functions.  {\it Glasgow Math J. } {\bf 30}, (1988), No. 1, 75-85, (1988).

\bibitem[Ni]{nier} F. Nier, A propos des fonctions theta et des r\'eseaux d?Abrikosov,
S\'eminaire Equations aux d\'eriv\'ees partielles (Polytechnique),  (2006-2007),  Talk no. 12.

\bibitem[OSP]{osp}  B. Osgood, R. Phillips, P. Sarnak, Extremals of Determinants of Laplacians, {\it J. 
Funct. Anal.} {\bf 80} (1988), 148--211.


\bibitem[PRN]{PRN} M. Petrache, S. Rota Nodari,  Equidistribution of jellium energy for Coulomb and Riesz Interactions, {\it Constr. Approx.}  {\bf 47}, (2018), 163-210.


\bibitem[PS]{ps} M. Petrache, S. Serfaty,  Next Order Asymptotics and Renormalized Energy for Riesz Interactions, {\it  J. Inst. Math. Jussieu} {\bf 16} (2017) No. 3, 501--569.


\bibitem[Ra]{radin} C. Radin, The ground states for soft disks,  {\it J. Stat. Phys. } {\bf 26} (1981), no. 2, 365--373.



\bibitem[Ran]{rankin} R. A. Rankin,
 A minimum problem for the Epstein zeta
function,. {\it Proc. Glasgow Math. Assoc, 1}  (1953), 149-158.



\bibitem[RoSt]{roncalstinga}L. Roncal, P. R. Stinga, Fractional Laplacian on the torus, {\it Comm. Contemp. Math.} {\bf 18.03} (2016), 1550033.


   
\bibitem[RNS]{rns} S. Rota Nodari, S. Serfaty, Renormalized energy equidistribution and local charge balance in 2D Coulomb systems, {\it Inter. Math. Res. Not.} {\bf 11}  (2015),  3035--3093.


\bibitem[RoSe]{rs} N. Rougerie, S. Serfaty, Higher Dimensional Coulomb Gases and  Renormalized Energy Functionals, {\it Comm. Pure Appl. Math} {\bf 69} (2016), 519--605.

\bibitem[SS1]{ssgl} E. Sandier, S. Serfaty, From the Ginzburg-Landau model to vortex lattice problems, {\it Comm. Math. Phys. } {\bf 313} (2012), 635--743. 

 \bibitem[SS2]{ss2}E. Sandier, S. Serfaty, 2D Coulomb Gases and the Renormalized Energy, {\it Annals of Proba}, {\bf 43}, no 4, (2015), 2026--2083.
 
 \bibitem[SaSt]{sarnak}
 P. Sarnak and A. Str\"ombergsson, Minima of Epstein's zeta function and heights
of flat tori, {\it Invent. Math. } {\bf165}  (2006), no. 1, 115--151. 
 
 \bibitem[S1]{ln} S. Serfaty, {\it Coulomb gases and Ginzburg-Landau vortices}, Zurich Lectures in Advanced Mathematics, 70, Eur. Math. Soc., 2015.
    

 \bibitem[S2]{sicm} S. Serfaty, Systems of points with Coulomb interactions,  Proceedings International Congress of Mathematicians, Rio de Janeiro, 2018.

\bibitem[Th]{theil} F. Theil, A proof of crystallization in two dimensions,  {\it Comm. Math. Phys.} {\bf 262} (2006), no. 1, 209--236.
\bibitem[Via]{via} M.  Viazovska, The sphere packing problem in dimension 8, {\it Ann. of Math.} {\bf 185}  (2) (2017), no. 3, 991--1015. 


\bibitem[Wi]{wigner} E. P. Wigner, On the interaction of electrons in metals, {\it Phys. Rev.} {\bf 46}, (1934), 1002--1011.

\end{thebibliography}
\end{document}